\documentclass{aa}
\pdfoutput=1
\usepackage[varg]{txfonts}
\usepackage{graphics,graphicx,enumitem}
\usepackage{hyperref}
\usepackage{natbib}
\usepackage{braket}
\bibpunct{(}{)}{;}{a}{}{,} 

\begin{document}	

\title{The coronagraphic Modal Wavefront Sensor: a hybrid focal-plane sensor for the high-contrast imaging of circumstellar environments}
\titlerunning{The coronagraphic Modal Wavefront Sensor}
\author{M.J. Wilby \inst{\ref{Leiden}} \and C.U. Keller \inst{\ref{Leiden}} \and F. Snik \inst{\ref{Leiden}} \and V. Korkiakoski \inst{\ref{ANU}} \and A.G.M. Pietrow \inst{\ref{Leiden}}}
\institute{Leiden Observatory, Leiden University, P.O. Box 9513, 2300 RA Leiden, The Netherlands\label{Leiden} \and Australian National University, Acton ACT 2601, Canberra, Australia\label{ANU} \\ \email{wilby@strw.leidenuniv.nl}}

\date{Pre-print online version: \today}

\abstract {The raw coronagraphic performance of current high-contrast imaging instruments is limited by the presence of a quasi-static speckle (QSS) background, resulting from instrumental non-common path errors (NCPEs). Rapid development of efficient speckle subtraction techniques in data reduction has enabled final contrasts of up to $10^{-6}$ to be obtained, however it remains preferable to eliminate the underlying NCPEs at the source. 
In this work we introduce the coronagraphic Modal Wavefront Sensor (cMWS), a new wavefront sensor suitable for real-time NCPE correction. This pupil-plane optic combines the apodizing phase plate coronagraph with a holographic modal wavefront sensor, to provide simultaneous coronagraphic imaging and focal-plane wavefront sensing using the science point spread function.
We first characterise the baseline performance of the cMWS via idealised closed-loop simulations, showing that the sensor successfully recovers diffraction-limited coronagraph performance over an effective dynamic range of $\pm 2.5 $~radians root-mean-square (RMS) wavefront error within 2-10 iterations. 
We then present the results of initial on-sky testing at the William Herschel Telescope, and demonstrate that the sensor is able to retrieve injected wavefront aberrations to an accuracy of 10~nm RMS under realistic seeing conditions. We also find that the cMWS is capable of real-time broadband measurement of atmospheric wavefront variance at a cadence of 50~Hz, across an uncorrected telescope sub-aperture. 
When combined with a suitable closed-loop adaptive optics system, the cMWS holds the potential to deliver an improvement in raw contrast of up to two orders of magnitude over the uncorrected QSS floor. Such a sensor would be eminently suitable for the direct imaging and spectroscopy of exoplanets with both existing and future instruments, including EPICS and METIS for the E-ELT.}


\maketitle

\section{Introduction}

	\subsection{Scientific motivation}
	
	Since the first direct image of a planetary mass companion around a nearby star was obtained in 2004 \citep{Chau:04}, the field of high-contrast imaging has undergone rapid development with the advent of advanced coronagraphic techniques \citep{Mawet:12} and eXtreme Adaptive Optics (XAO) systems (eg. \citealp{Sauvage:10}). This progress continues with the recent first light science of the high-contrast imaging instruments GPI \citep{Mac:14}, SPHERE \citep{Beuzit:08} and ScExAO \citep{Jova:15}, which are detecting and characterising young gaseous exoplanets with ever lower masses approaching that of Jupiter \citep{Mac:15, Bonne:15} and comprehensively studying planet-disk interactions and the planet formation process (eg. \citealp{Avenhaus:14,Benisty:15}). Such work is also informing the design parameters of the next generation of ground-based ELT-class instruments which aim to characterise rocky exoplanets in the habitable zones of nearby stars. This challenging goal requires final contrast ratios of better than $10^{-7}$ at inner-working angles of the order 10 milli-arcseconds \cite{Guyon:12}, starting with planets orbiting M-dwarf host stars such as the newly discovered Proxima Centauri b \citep{Anglada:16}. The expected limit on achievable raw imaging contrast with ground-based, coronagraph-enabled XAO systems is of the order of $10^{-5}$ for large field-of-view starlight suppression regions \citep{Kasper:10,Guyon:12}, hence this must be combined with complementary high-contrast techniques such as polarimetric differential imaging \citep{Keller:10,Perrin:15} and high-dispersion spectroscopy \citep{Snellen:15}, which are already expanding the toolkit of the exoplanet imaging community.
	
	Of the diverse approaches to high-contrast imaging and specifically coronagraphy, the Apodizing Phase Plate (APP) coronagraph \citep{Codona:06,Kenworthy:10b,Quanz:10} is of particular relevance to this paper. This technique uses a pupil-plane phase mask to modify the point-spread function (PSF) of the instrument, thereby using destructive interference to create a “dark hole” in the diffracted stellar halo at the location of the planet. This approach makes the APP an extremely versatile coronagraph, allowing simultaneous coronagraphic observation of multiple targets in the same field, providing insensitivity to tip-tilt errors, and reducing the pointing tolerances on chopping offsets required for accurate background subtraction at the near-infrared wavelengths most favourable for observation of young, thermally luminous exoplanets.
	The recent development of the vector-Apodizing Phase Plate (vAPP, \citealp{Otten:14}), which provides simultaneous $360^\circ$ coverage around the host star by using circular polarisation beam-splitting to create duplicate copies of the classical APP pattern,	has also accompanied significant gains in inner-working angle, with a vAPP operating at radial separations of 1.2-6~$\lambda/D$ (where $\lambda$ is the observing wavelength and $D$ is the telescope diameter) installed and available for science observations in MagAO \citep{Morzinski:14} at the Magellan Clay Telescope (Otten et al., submitted).
	
	These ground-based, XAO-corrected high-contrast imagers are limited by ever-present Non-Common Path Errors (NCPEs); these wavefront aberrations arise due to the presence of differential optics between the AO wavefront sensor and the science focal plane, which may be influenced by slow thermal or mechanical fluctuations. The resulting focal-plane quasi-static speckle (QSS) field is coherent on timescales of minutes to hours, and limits the raw performance of most coronagraphs to $10^{-4}-10^{-5}$ in contrast, defined here as the $5\sigma$ companion detectability limit, over an entire observation period \citep{Martinez:12}. Advanced observation and data reduction algorithms such as the Locally Optimised Combination of Images (LOCI) \citep{Lafren:07} and Principle Component Analysis \citep{Soummer:12,Amara:12} have been used to surpass this limit and achieve detection thresholds of $10^{-6}$ at separations larger than 7~$\lambda/D$ with SPHERE and GPI (\citealp{Zurlo:16,Mac:14}). However, due to the impact of quasi-static speckles on the ultimate photon noise limit, in addition to ongoing uncertainties surrounding the influence of post-observation NCPE suppression algorithms on the derived properties of subsequently detected companions (e.g. \citealp{Marois:10}), it remains preferable to correct these non-common path errors in real time and thereby return coronagraphic performance to the diffraction-limited regime.

	The complete elimination of NCPEs ultimately relies on the principle of focal-plane wavefront sensing; only by using the sicence camera as a sensor can the AO loop have a truly common path with observations. Existing focal-plane wavefront reconstruction techniques use artificially induced phase diversity \citep{Keller:12,Kork:13} or properties of the speckle field itself \citep{Codona:13} to overcome the degeneracies associated with a loss of wavefront spatial resolution and incomplete knowledge of the focal-plane electric field. Although there have been some successful on-sky demonstrations of these techniques (eg. \citealp{Martin:14}), factors such as computational complexity, invasive modification of the science PSF, and limited dynamic or chromatic range mean that such reconstruction methods have not yet been widely adopted for science observations. To avoid these limitations many high-contrast imaging instruments instead perform periodic offline NCPE calibrations, such as the COFFEE coronagraphic phase diversity algorithm proposed for use in SPHERE \citep{Sauvage:11}, at the cost of temporal resolution and the loss of simultaneity with science observations.

	There is therefore an ongoing drive to develop a coronagraphic focal-plane wavefront sensor which is able to operate in parallel with science imaging in a non-invasive manner, and provide unbiased real-time compensation of the low spatial frequency NCPEs which correspond to small angular separations in the observed stellar image.
	
	\subsection{Holographic optics for focal-plane wavefront sensing}

	The use of computer-generated holograms as a method of focal-plane wavefront sensing has been extensively explored in the literature, with specific focus on applications in confocal microscopy \citep{Neil:00,Booth:03} and laser collimation \citep{Changhai:11}. This approach is used to generate secondary PSF copies in the science focal plane, which are spatially separated from the main science PSF to avoid mutual interference. In the so-called Holographic Modal Wavefront Sensor (HMWS) these wavefront-sensing PSFs are artificially biased with a set of chosen aberration modes drawn from a suitable basis set (for example the Zernike modes), such that the Strehl ratio of each PSF copy responds linearly to the corresponding aberration mode present in the input wavefront. In this way the sensor performs a modal decomposition of the incoming wavefront into the chosen basis, which may be reconstructed in real time with the intensity measurement of two focal-plane photometric apertures per mode.
	
	\begin{figure*}
	\centering
	\includegraphics[width=0.9\textwidth]{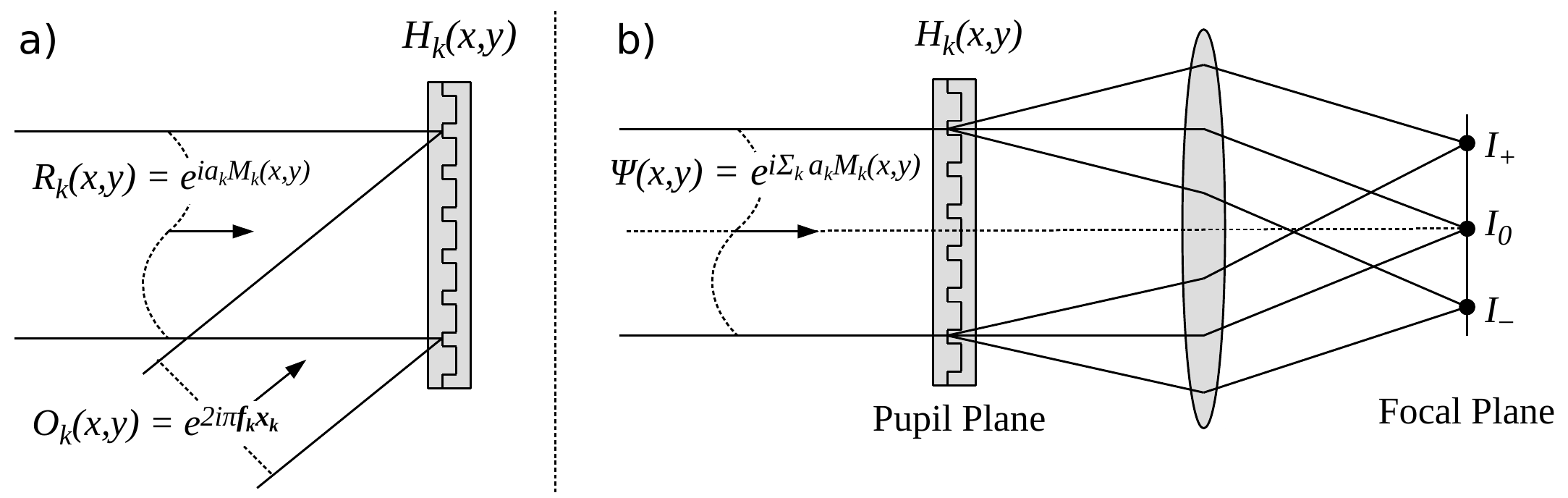}
	\caption{Diagram showing the principle of HMWS operation. a) Visual representation of the creation process of a single-mode computer-generated hologram, by analogy with optical exposure. b) Operation of a single-mode hologram in the presence of an aberrating wavefront. Figure adapted from \cite{Dong:12}.}
	\label{Fig:1_HMWS_diagram}
	\end{figure*}
	
	A modal approach to wavefront sensing has multiple advantages over traditional wavefront sensors as well as other focal-plane wavefront sensing techniques, most notably in terms of reduced computational complexity, the fact that the resolution of the reconstructed wavefront is not limited to the spatial resolution of the sensor's pupil element as with a Shack-Hartmann sensor, and that a modal wavefront is simple to implement on many current deformable elements. 
	For the science case of high-contrast imaging of exoplanets and circumstellar environments, the HMWS should operate simultaneously with a coronagraph in the science focal plane, to directly retrieve the aberrations that are seen by the starlight suppression system in the instrument.
	We therefore consider here the promising combination of the HMWS with the APP coronagraph: for the purposes of this paper we shall refer to the resulting optic as the coronagraphic Modal Wavefront Sensor (cMWS). This hybrid approach can be easily implemented since both concepts are phase-only pupil plane optics, which may be easily multiplexed into a single physical element. The HMWS is however not limited to use with pupil plane phase-only coronagraphs, provided that the hologram is positioned upstream of any focal-plane masking elements in order to transmit the central diffraction core of all holographic PSF copies.
		
	\subsection{Content of paper}
	This paper is divided into the following sections: in Sec.~\ref{Sec:Theor} we summarise the underlying mathematics behind holographic modal wavefront sensing, and present the critical factors which must be considered when multiplexing the HMWS with an APP coronagraph. Sec.~\ref{Sec:Sims} shows the results of idealised closed-loop simulations and outlines the baseline performance of the sensor for the case of a clear circular aperture. Sec.~\ref{Sec: WHT Results} presents results from the first on-sky implementation of a cMWS sensor at the William Herschel Telescope (WHT) located in La Palma, Spain, including a demonstration of sensitivity to both static and dynamic wavefront errors. In Sec.~\ref{Sec: Conclusion} we draw final conclusions and present goals for ongoing and future work.

\section{Theory}
\label{Sec:Theor}
	
	\subsection{The Holographic Modal Wavefront Sensor}
	
	The principle of the HMWS relies on the fact that the phase component $\phi(x,y)$ of an arbitrary wavefront may be decomposed into coefficients of a chosen 2D mode basis describing the telescope aperture, for which the complex electric field $\Psi(x,y)$ may be written as
	
	\begin{equation}	
	\Psi(x,y) = A(x,y)e^{i\phi (x,y)} = A(x,y)e^{i\sum\limits_j a_jM_j(x,y)},
	\label{Eq:Mode_basis}
	\end{equation}
	
	where $A(x,y)$ is the telescope aperture function, $M_j(x,y)$ is some complete, and ideally orthonormal, mode basis with RMS coefficients $a_j$ (in radians) and $x,y$ are coordinates in the pupil plane. In this paper we focus exclusively on phase-only aberrations as these are simpler to implement and correct for, and dominate the total wavefront error in almost all practical cases.
	
	In order to provide full phase-aberration information in a single focal-plane intensity image, the sensor uses a computer-generated holographic element to perform an instantaneous modal decomposition and extract the set of coefficients $a_j$, albeit up to a truncated mode order. It is then trivial to reconstruct the wavefront using the set of template modes using Eq.~\ref{Eq:Mode_basis}, which may then be passed to an adaptive optics system for correction either as a direct command or via the adjustment of reference slope offsets.
	
	\subsubsection{Generating holograms}
	\label{Sec:HoloGen}

	The purpose of the hologram in a HMWS is twofold: firstly it creates secondary PSF copies which are spatially separated from the zero-order PSF in the science focal plane. Secondly, it adds an artificial bias wavefront independently to each of these PSF copies, such that each responds differently to the input wavefront $\Psi$. This can therefore be thought of as a system of 2N simultaneous phase diversities chosen to span the desired mode basis, but instead of the normal approach to focal-plane phase diversity reconstruction (which typically uses only one diversity and the intensities of all pixels in the PSF), the modal content of the wavefront is extracted in a more direct fashion by measuring only the relative core intensities of all PSF copies.

	As illustrated in Fig.~\ref{Fig:1_HMWS_diagram}a, the holographic element is constructed numerically from two independent components which perform the functions described above. Adopting the notation of \cite{Dong:12}, the reference wave $R_k$
	
	\begin{equation}
	R_k(x,y)=e^{ib_kM_k(x,y)}
	\label{Eq:R}
	\end{equation}
	
	contains a single bias mode $M_k$ with an RMS aberration strength (in radians) set by the bias strength $b_k$. The object wave $O_k$ is given by
	
	\begin{equation}
	O_k(x,y) = e^{2i\pi(f_{kx}x+f_{ky}y)} ,
	\label{Eq:O}
	\end{equation}
	
	 where the spatial frequencies $f_{kx,y} = x_k',y_k'/f\lambda$ specify the desired tilted plane wave and thus the coordinates $(x_k',y_k')$ in the focal plane. The holographic phase pattern $H_k(x,y)$ for this particular mode is then the interferogram between these two waves,
	
	\begin{align}
	H_k(x,y) &= \left|O_k(x,y) + R_k(x,y)\right|^2 \\
	&= \left|O_k\right|^2 + \left|R_k\right|^2 + O_k^*R_k + O_kR_k^* \\
	&= 2 + 2 \mathbb{Re}\left[O_k^*R_k\right] ,
	\label{Eq:H_k}
	\end{align}
	
	 where $^*$ is the complex conjugate operator and $\mathbb{Re}\left[\right]$ denotes the real component of the complex argument.
	
	It follows from this that the two conjugate terms naturally result in the creation of two wavefront sensing spots which may be treated as the $\pm 1$ orders of a diffraction grating, containing equal and opposite bias aberrations $\pm b_k$ . The first two terms in Eq.~5 are equal to unity and are discarded such that $\langle H_k \rangle = 0$. 
	The behaviour of this hologram in the presence of an aberrated wavefront $\Psi$ is shown graphically in Fig.~\ref{Fig:1_HMWS_diagram}b. The total focal-plane intensity is then given by $I = \left|\mathcal{F}\left[H_k\Psi\right]\right|^2$, where $\mathcal{F}$ is the Fourier Transform operator in the Fraunhofer diffraction regime. Following from this and Eqs.~\ref{Eq:Mode_basis},~\ref{Eq:R},~\ref{Eq:O}~and~\ref{Eq:H_k}, the local intensity distribution $I_{k\pm}$ of the pair of biased PSF copies is given by
	
	\begin{align}
	\begin{split}
	I_{k\pm}(x',y')&=\underbrace{\delta({\bf x'} \pm {\bf x_k'})}_{\text{Carrier Frequency}} \ast~ \underbrace{{\Big|}\mathcal{F}{\Big[}A(x,y){\Big]}}_{\text{Telescope PSF}}{\Big|}^2\\
	&\ast~  
	{\Big|}\mathcal{F}{\Big[}\underbrace{e^{i(a_k\pm b_k)M_k(x,y)}}_{\text{Wavefront Bias}} \underbrace{e^{i\sum_{j \neq k}a_jM_j(x,y)}}_{\text{Inter-Modal Crosstalk}}{\Big]}{\Big|}^2 ,
	\label{Eq:I_k}
	\end{split}
	\end{align}
	
	where $I_{k\pm}$ correspond to the positively and negatively biased wavefront sensing spots respectively, and $a_j$ is the RMS error present in the incident wavefront corresponding to mode $M_j$. The term $\delta({\bf x'})$ is the 2D delta function, with focal-plane coordinates ${\bf x_k'} = (x_k',y_k')$ deriving directly from the frequency of the carrier wave $O_k$. The second term encompasses the desired sensor response to the aberrated wavefront, with net aberration $a_k\pm b_k$. The final term represents a fundamental source of inter-modal crosstalk as a convolution with all other modes present in the input wavefront, which acts equally on both $I_{k\pm}$; see Section~\ref{Sec: WF_reconst} for a full discussion of the impact of this term. 
	
	An arbitrary number of holograms may be multiplexed into a single element, allowing the generation of multiple pairs of independently biased PSF copies and hence the simultaneous coverage of many wavefront modes. For simplicity of implementation we now create a phase-only hologram $\phi_h(x,y)$ by taking the argument of the multiplexed hologram

	\begin{equation}
	\phi_h(x,y) = \frac{s}{\pi} {\rm arg}\left[H(x,y)\right] = \frac{s}{\pi} {\rm arg}\left[\sum_k^N H_k(x,y))\right] ,
	\label{Eq:Holo}
	\end{equation}	
	
	which is by definition binary as all $H_k$ are real from Eq.~\ref{Eq:H_k}, and is normalised to have a grating amplitude of $s$ radians. Scaling down the amplitude from (0, $\pi$) allows direct control over the fractional transmission to the zeroth order, which forms the science PSF. It is assumed here that the holographic PSF copies are located sufficiently far from each other and the zeroth-order in the focal plane that there is negligible overlap; if this is not the case there will be additional inter-modal crosstalk in the sensor response due to mutual interference, which is independent of that arising from the final term of Eq.~\ref{Eq:I_k}.
	
	The optimal positioning of WFS copies for minimal inter-modal crosstalk is a significant optimisation problem in itself, which will be investigated in future work. As a rule of thumb, each spot should be positioned at least 5-6 $\lambda/D$ from not only all other first order PSF copies, but also from the locations of all corresponding higher-order diffraction copies and cross-terms; see the treatment in \cite{Changhai:11} for full details. In the general case this requires the computation of an appropriate non-redundant pattern, which is outside the scope of this paper, however a circular or "sawtooth" geometry (the latter is shown in Fig.~\ref{Fig:2_HMWS_simImg}) was found to be a suitable alternative geometry for the prototype cMWS.

	\subsubsection{Sensor response}
	
	Following the approach of \cite{Booth:03} it is possible to approximate the sensor response for $a_k \ll b_k$ as the Taylor expansion of Eq.~\ref{Eq:I_k} about $a_k=0$, where the on-axis intensity of each PSF copy can this way be expressed as

	\begin{equation}
	I_{k\pm} = I_0\left[f(b_k)\pm a_kf'(b_k) + \frac{a_k^2}{2}f''(b_k)+\mathcal{O}(a_k^3)\right]
	\label{Eq:Taylor}
	\end{equation}	
	
	where $I_0$ is a multiplicative factor proportional to total spot intensity and $f(b_k)=\left|1/\pi \iint e^{ib_kM_k(x,y)}dxdy\right|^2$ is the Fourier integral for an on-axis detector of infinitesimal size. Throughout this paper we adopt the normalised intensity difference between spot pairs as the metric for sensor measurement, equivalent to the "Type B" sensor of \cite{Booth:03}. In this case, the sensor response per mode $I_k$ is given by
	
	\begin{equation}	
	I_k= \frac{I_{k+}-I_{k-}}{I_{k+}+I_{k-}} = \frac{2a_kf'(b_k)+\mathcal{O}(a_k^3)}{2f(b_k)+a_k^2f''(b_k) + \mathcal{O}(a_k^5)} .
	\label{Eq:TypeB}
	\end{equation}
	
	If $b_k$ can be chosen such that $f''(b_k) = 0$, this expression becomes linear to 3rd order: for a Zernike basis \cite{Booth:03} find that this occurs for values of $\langle b_k \rangle = 1.1~{\rm rad}$, while values of $\langle b_k \rangle = 0.7~{\rm rad}$ resulted in maximal sensitivity; we adopt the latter value throughout this work. In principle the improved "Type C" sensor also suggested by \cite{Booth:03}, which uses the metric $I_k = (I_{k+} - I_{k-})/(I_{k+} + \gamma I_0 + I_{k-})$, can yield further improved linearity and suppression of intermodal crosstalk, however the inclusion of additional measurement requirements of an unbiased PSF copy $I_0$ and free parameter $\gamma$ (which must be determined empirically) make this unnecessary for use in a first implementation of the sensor.

	\subsubsection{Wavefront reconstruction}
	\label{Sec: WF_reconst}
	
\begin{figure}
	\centering
	\includegraphics[width=0.49\textwidth]{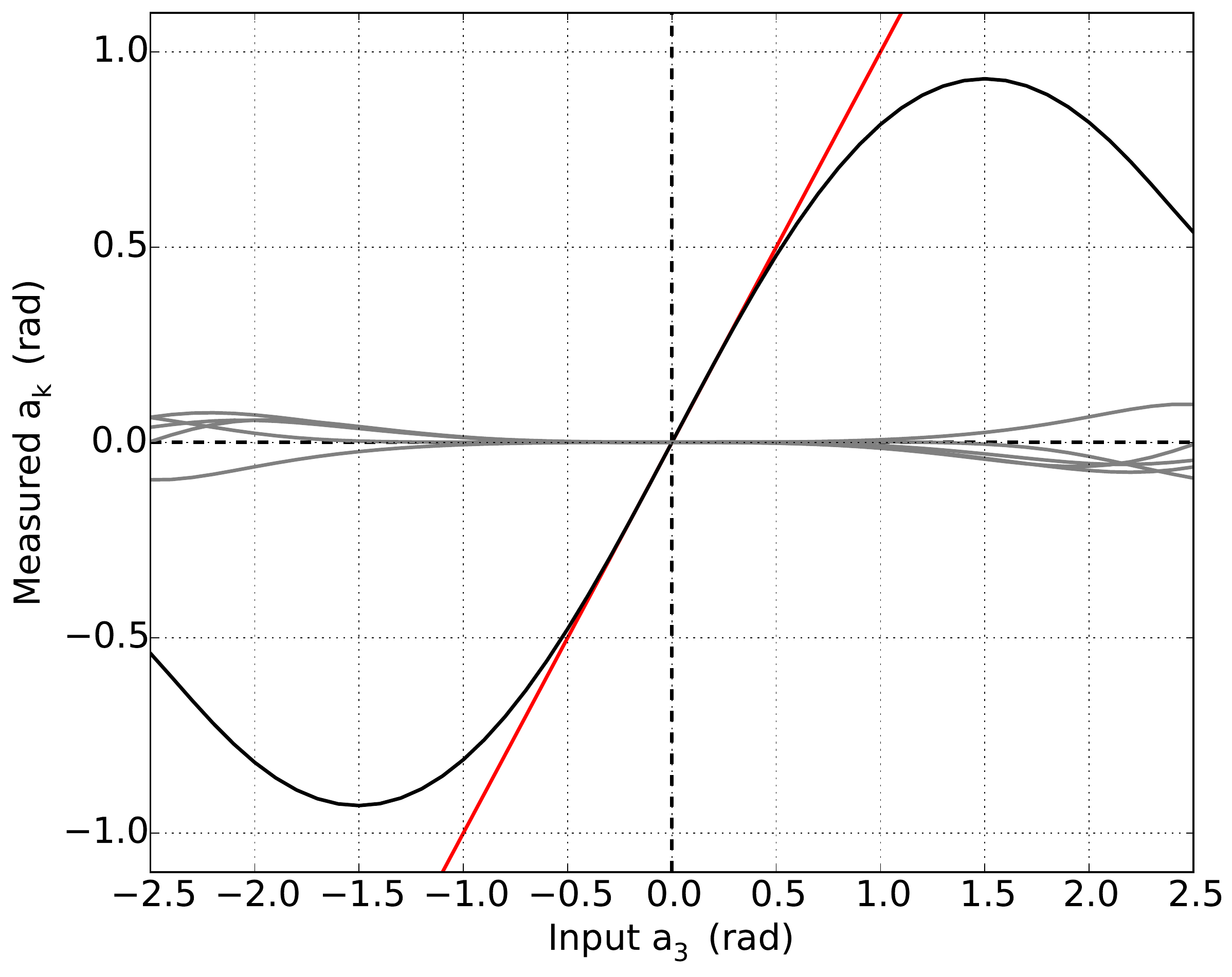}
	\caption{Response curve of a 6-mode Zernike HMWS ($b_k = 0.7$~rad) to defocus error $a_3Z_3$ (all $a_{k\neq 3}=0$), using photometric apertures of radius $r=1.22\lambda/D$. The diagonal line (red) denotes a perfect sensor with 1:1 correspondence, which is achieved by the HMWS for $\left|a_3\right| \lesssim 0.5 $. The response of the remaining sensor modes ($Z_4$-$Z_8$, grey) are well constrained about zero over the linear sensing regime, with residual nonlinear inter-modal crosstalk behaviour manifesting for $\left|a_3\right| \gtrsim 1 $.}
	\label{Fig:2c_Rcurve}
	\end{figure}

	Final estimates of the mode coefficients $a_k$ of the incoming wavefront must then be obtained by calibrating intensity measurements with a cMWS response matrix ${\bf \hat{G}}$, which provides the nominal scaling factors between $I_k$ and $a_k$ but is also capable of providing a first-order correction for inter-modal crosstalk via its off-diagonal terms. This matrix is formed from the gradients of the characteristic response curves $I_i(a_j)$ response curves about $I = 0$,
	
	\begin{equation}
	G_{ij} = \left.\frac{\delta I_{i}}{\delta a_j}\right|_{I=0} , 
	\label{Eq:S_ij}
	\end{equation}
	where $G_{ij}$ is the response of sensor mode $i$ to input wavefront error of mode $j$. The solution for the set of mode coefficients ${\bf a}$ of the incoming wavefront is then
	\begin{equation}
	{\bf a} = {\bf \hat{G}}^{-1}{\bf I} , 
	\label{Eq: CT_Calib}
	\end{equation}
	
	where ${\bf a}$ and $\bf{ I}$ are the column vectors comprising sensor response $I_k$ and the corresponding wavefront coefficient estimates $a_k$ respectively. Note that the standard multiplicative inverse ${\bf \hat{G}}^{-1}$ of the interaction matrix is used here, since the interaction matrix is square, highly diagonal and with on-diagonal elements defined so as to have the same sign. It is therefore extremely unlikely that this matrix is degenerate and thus non-invertible, but in such a case the Moore-Penrose pseudo-inverse ${\bf \hat{G}}^{+}$ (see e.g. \citealp{Barata:12}) may be used as an alternative.
	Fig.~\ref{Fig:2c_Rcurve} shows an illustrative response curve to which this calibration has been applied, showing that the sensor response is linear over the range $\left|a_k\right| \lesssim b_k $ with negligible inter-modal crosstalk, beyond which wavefront error is increasingly underestimated as the main assumption of Eq.~\ref{Eq:Taylor} begins to break down. A turnover in sensitivity occurs at the point $a_k = 2b_k$ since beyond this the input wavefront error dominates over the differential bias $\pm b_k$.
		
		In addition to calibrating sensor measurements to physical units, ${\bf \hat{G}}$ also performs a linear correction for inter-modal crosstalk; this allows the knowledge of the responses of all other spot pairs to be used to infer the correct mode measurement of one particular pair. As denoted by the final term in Eq.\ref{Eq:I_k}, this effect occurs via a convolution of the WFS spot $I_k$ with all remaining wavefront aberrations $M_{j\neq k}$ present in the input wavefront. This effect was neglected in the previous section as the convolution term is reduced to a constant multiplicative factor under the on-axis assumption, factoring out in Eq.~\ref{Eq:TypeB}. The theoretical response matrix for any set of orthogonal modes is therefore diagonally dominated and sparse \citep{Booth:03}, but in practice many factors such as use of photometric apertures of non-zeros size, alignment errors or overlap with the wings of other PSF copies or the zeroth order, may result in significantly elevated crosstalk behaviour. 
		
		Empirical determination of a full response matrix for each cMWS design is therefore the most robust method of compensating these effects to first order. This process is straightforward and once automated takes only a few minutes to perform: each column of the interaction matrix requires a minimum of two measurements of the normalised intensity vector ${\bf I}$, each for different known coefficents $a_k$ of the corresponding input wavefront mode applied on the corrective element, in order to fit the gradients of each response curve. This procedure is in principle required only once for any given instrument configuration, however performing regular re-calibration before each observation night is feasible and allows the elimination of slow drifts in actuator response or instrument alignment quality.
	
	\subsection{Combination with an Apodizing Phase Plate coronagraph}
	\label{Sec: APP+CGH}
	
	The APP is an optimal coronagraph for use in the cMWS as not only is it a pupil plane phase only optic and thus simple to multiplex with the HMWS, but is also preserves an Airy-like PSF core required for production of holographic copies. By contrast, focal-plane or hybrid coronagraphs would require the hologram to be located in pupil upstream of the focal-plane mask in order to create the off-axis PSF copies before rejection of on-axis stellar light occurs.
	The resulting optic may be implemented using the same techniques as for the APP; as either a transmissive optic such as a turned glass phase plate \citep{Kenworthy:10} or achromatic liquid crystal retarder \citep{Snik:12}, or via a phase-apodizing Spatial Light Modulator (SLM) \citep{Otten:14}.	
	
	Consider now the combination of the HMWS presented above with an APP coronagraph into a single optic such that the modification to the complex wavefront $\Psi_{cMWS}(x,y)$ may be described as
	
	\begin{equation}	
	\Psi_{cMWS}(x,y) = A(x,y)e^{i\left[\phi_c (x,y)+\phi_h (x,y)\right]} ,
	\label{Eq:APP_HMWS_WF}
	\end{equation}	
	
	where $\phi_c(x,y)$ and $\phi_h(x,y)$ correspond to the coronagraph and normalised hologram (Eq.~\ref{Eq:Holo}) phase patterns respectively.
	
\begin{figure*}
	\centering
	\includegraphics[width=0.95\textwidth]{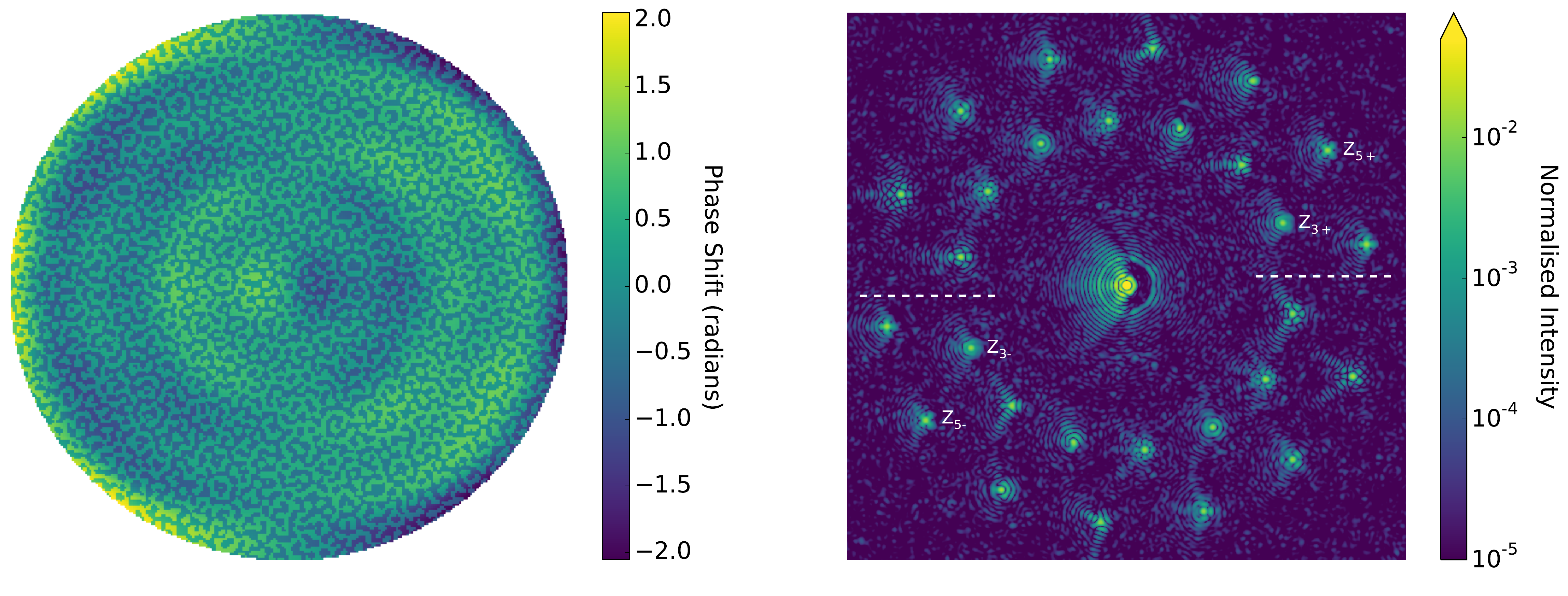}
	\caption{Simulation of a 14-mode Zernike cMWS (modes $Z_3-Z_{16}$) combined with an APP optimised for a $10^{-6}$ dark hole  with a $180^\circ$ opening angle. {\bf Left:} Multiplexed pupil-plane phase design $\Psi_{cMWS}$ containing the high spatial frequency HMWS binary grating overlaid on the smoother APP design. {\bf Right:} Corresponding focal-plane PSF: positively and negatively biased PSF copies are located in the top and bottom half of the image respectively, separated by the white dashed lines. Two example pairs are labelled (defocus, $Z_3$ and $45^\circ$ astigmatism, $Z_5$), illustrating the symmetry of the $\pm1$ orders about the zeroth order PSF.}
	\label{Fig:2_HMWS_simImg}
	\end{figure*}

	Fig~\ref{Fig:2_HMWS_simImg} shows the simulated pupil optic and corresponding PSF of a cMWS coded for the 14 lowest order nontrivial Zernike modes, including an APP with a 180 degree dark hole extending from $2.7 - 6 \lambda/D$, generated using a Gerchberg-Saxton style iterative optimisation algorithm. The hologram pattern is seen in the pupil as an irregular binary grating overlaid on top of the smooth phase variations of the APP. The wavefront sensing spots can clearly be seen surrounding the dominant central science PSF, with the PSF of each copy formed by the convolution of the characteristic Zernike mode PSF with that of the APP. For illustration purposes a grating amplitude of $s=\pi/2$ here results in an average normalsed intensity difference of -1.8dex between the peak flux of each WFS copy and the zeroth order PSF, with an effective transmission to the science PSF of $50\%$. It is however possible to operate the sensor with significantly fainter PSF copies in practice, making $80-90\%$ transmission achievable with respect to the APP alone.
	
	 \subsection{Impact of multiplexing on mutual performance}	
	 
\begin{figure}
	\centering
	\includegraphics[width=0.49\textwidth]{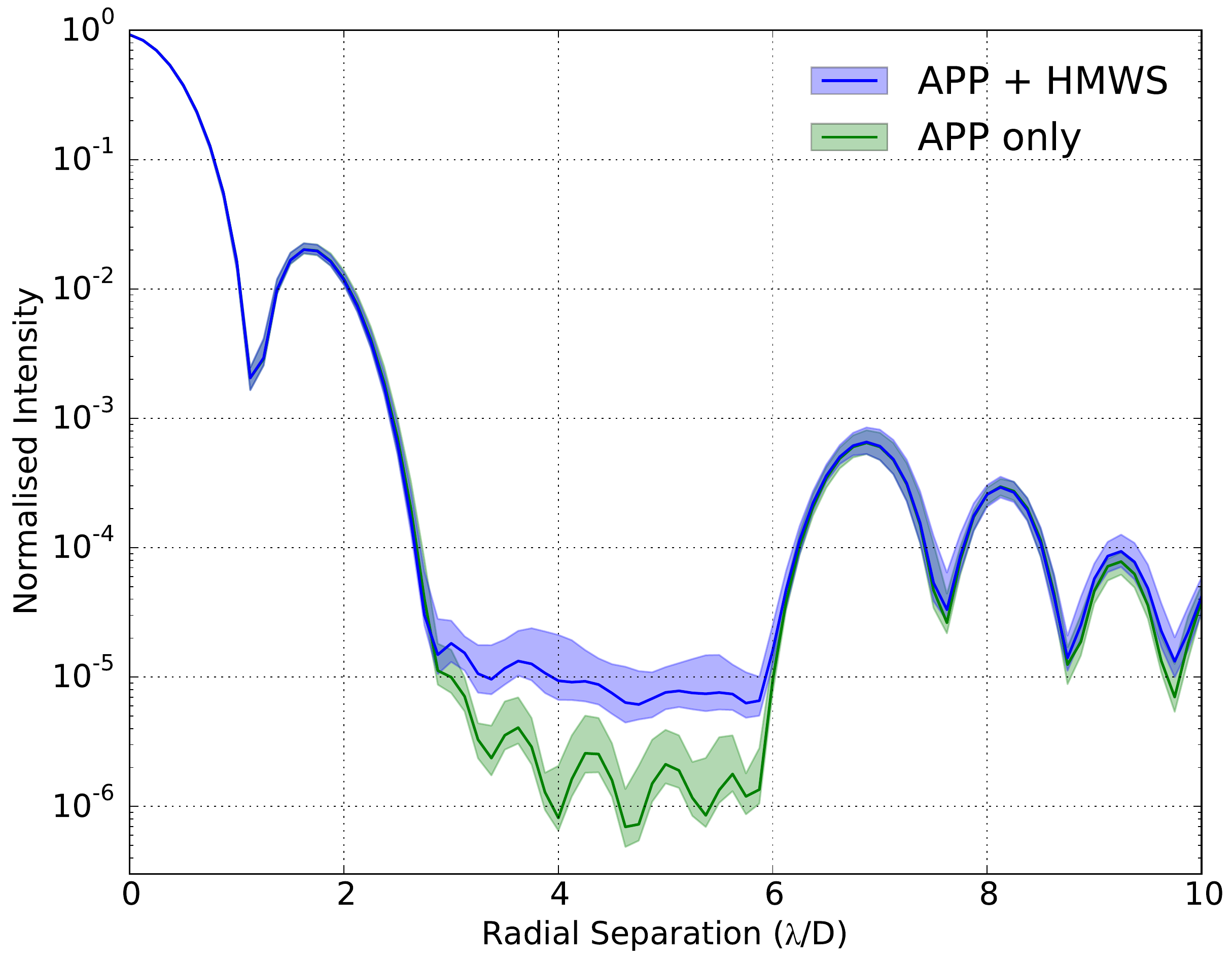}
	\caption{Contrast curves for the APP coronagraph presented in Fig.~\ref{Fig:2_HMWS_simImg}, with (blue) and without (green) the HMWS hologram. The shaded regions denote the 1-sigma variance limit of residual structure at each radius, azimuthally averaged over a 170 degree region corresponding to the dark hole contrast floor.}
	\label{Fig:2.5_DH_filling}
	\end{figure}

	As the zeroth order PSF may be considered a 'leakage' term of the binary hologram grating, the APP pattern is in principle independent of all wavefront biases which appear in the $\pm 1$ diffraction orders. However there are two notable effects which must be considered when multiplexing these two optics, the first of which is that any stray light scattered by the HMWS will fill in the coronagraphic dark hole. As shown in Fig.~\ref{Fig:2.5_DH_filling}, it was found that the binary holograms generate a near-constant intensity scattered background at a mean normalised intensity of the order of $10^{-5}$, irrespective of the specific HMWS or APP designs used. This behaviour is due to the loss of information associated with creating a binary optic from the full complex hologram in Eq.~\ref{Eq:Holo}. Although a limiting dark hole depth of $10^{-5}$ remains sufficient for a first prototype, it would be possible to compensate for this effect by re-optimising the APP in the presence of the scattered background. 
		
		The second effect of the multiplexing process is that, as can be seen from Eq.~\ref{Eq:APP_HMWS_WF}, the APP phase pattern introduces a set of static wavefront errors which must be disregarded by the HMWS. This can be achieved by adding static reference slope offsets to to Eq.~\ref{Eq: CT_Calib} in a similar manner to existing NCPE correction routines (eg. \citealp{Sauvage:11}), such that
	
	\begin{equation}
	{\bf a} = {\bf \hat{G}}^{-1}{\bf I} - {\bf a_c}
	\label{Eq: APP_CTcorr}
	\end{equation}
	
	where ${\bf a_c}$ is the set of coefficients of $\phi_c$ in the sensing mode basis. This must be determined independently from ${\bf \hat{G}}$ to avoid degeneracy with static instrumental wavefront errors, either by projecting the APP onto the sensing mode basis $ a_{c,i} = \phi_c(x,y) \cdot M_i(x,y)$, or by comparison with calibration data containing only the non-multiplexed HMWS component.
	
	\subsection{Impact of structured telescope apertures}
	\label{Sec: AperFns}
	
	It is important to note that throughout this paper the cMWS is  evaluated for use with an un-obscured circular aperture, however it must also be applicable to more complicated amplitude profiles featuring central obscurations, support spiders, and mirror segmentation. If no modifications to the cMWS design are made, any aperture modifications will degrade the orthogonality of the chosen mode basis and thus lead to increased inter-modal crosstalk. 
	Fortunately this is not considered to be a limiting factor of the cMWS, as the effect can be effectively eliminated by performing a re-orthogonalisation of the chosen mode basis using the known aperture function, for example by using a simple Gram–Schmidt procedure (see e.g. \citealp{cheney:09}). This approach has now been verified during a more recent observing campaign at the WHT, the details of which will be the subject of a future work.
	In the case where the aperture function contains significant structures which are not azimuthally uniform, such as especially thick telescope spiders or mirror segmentation gaps, this procedure will be most effective when operated in a pupil-stabilised observation mode. This will allow the telescope aperture function to remain consistent with that of the re-orthogonalised sensing basis for the duration of each observation, however it was seen that in the case of the WHT pupil the 1.2~m circular central obscuration was in practice the only significant structure.

It is in principle also possible to develop the cMWS as a co-phasing sensor for segmented mirrors, for which the ideal sensing basis would instead consist of differential piston, tip and tilt modes which directly match the degrees of freedom of each individual mirror segment.
	That being said, the cMWS is not an ideal choice of sensor for co-phasing large future segmented telescopes such as the European Extremely Large Telescope (E-ELT) or the Thirty Meter Telescope (TMT), principally because the sensing basis would need to consist of an unreasonably large number of modes (2394 in the case of the E-ELT) in order to fully describe all possible phasing errors. While it may be possible to achieve this by sequentially correcting with multiple cMWS designs each containing a subset of the possible modes, such applications are much more suited to telescopes with significantly fewer mirrors where the calibration may be performed for all segments simultaneously, such as the W.M. Keck Observatory in Hawaii, or the Giant Magellan Telescope (GMT).

\section{Idealised performance simulations}
\label{Sec:Sims}
		
\begin{figure*}
		\centering
		\includegraphics[width=0.83\textwidth]{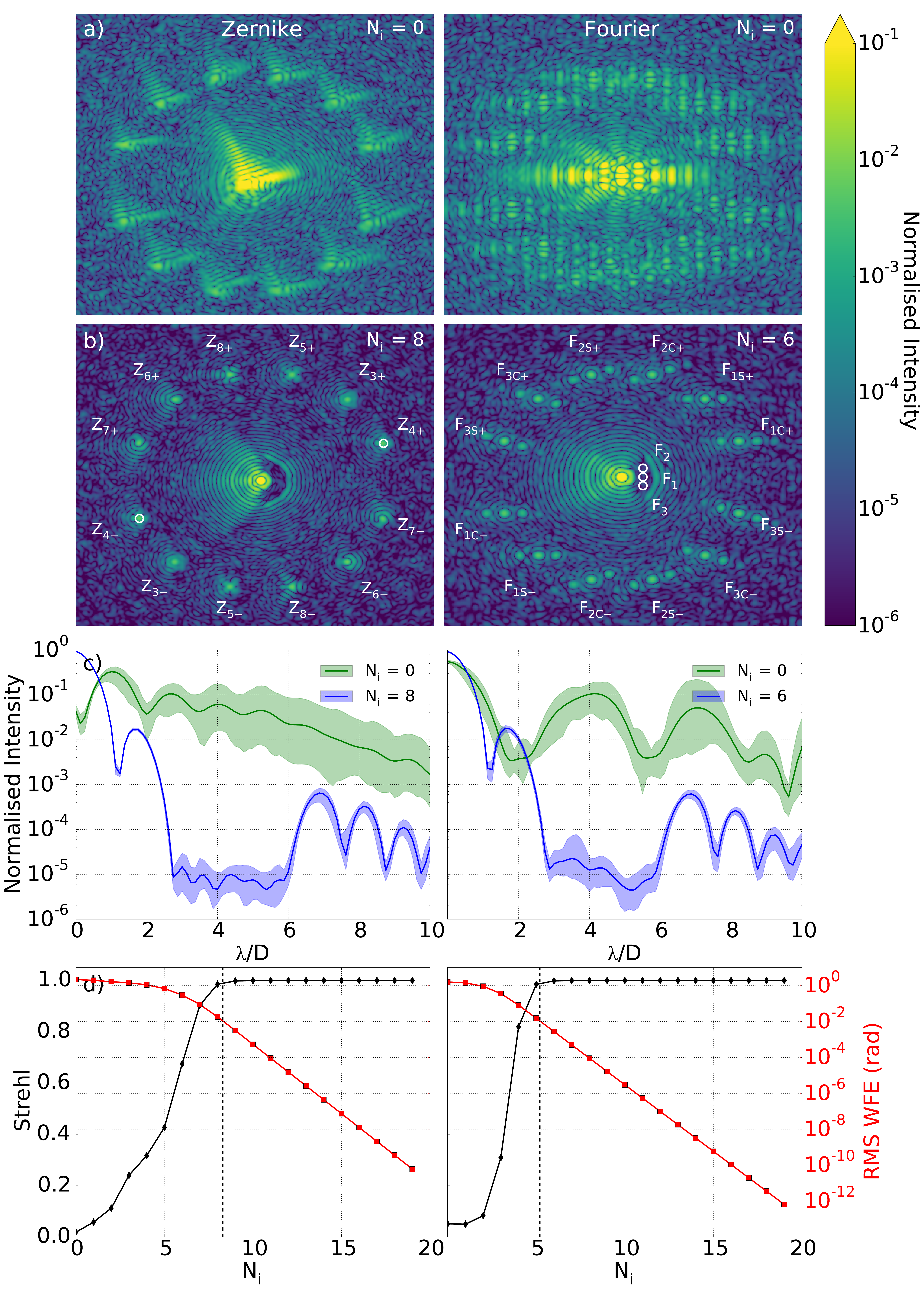}
		\caption{Example closed-loop performance for a 6-mode Zernike (left column) and Fourier (right column) mode cMWS for $g=0.8$. {\bf a)} $N_i = 0$ aberrated PSFs, with $a_k=1.0$~radians RMS per mode. {\bf b)} Diffraction-limited PSF after $N_i$ closed-loop iterations required to achieve convergence. PSF copies corresponding to each mode bias are labelled $Z_{n\pm}$ for Zernike modes and $F_{nS/C\pm}$ for Fourier modes, where the index $S/C$ denote the sine and cosine mode phases respectively, and mode number corresponds to the circled regions of influence in the APP dark hole. The white circles overlaid on the $Z_4\pm$ modes indicated the $r=1.22\lambda /D$ region of interest used for wavefront measurement. Note also the differing angular extent of each APP, which cover $180^\circ$ and $90^\circ$ for the Zernike/Fourier designs respectively. {\bf c)} Azimuthally-averaged residual intensity plots corresponding to the PSFs of panels a) (green) and b) (blue); shaded regions denote $1\sigma$ variance averaged over the APP dark hole. {\bf d)} Science PSF Strehl ratio (black diamonds) and residual RMS wavefront error (red squares) as a function of iteration number $N_i$. Vertical dashed lines indicate the point of convergence.}
		
		\label{Fig:CL_sims}
		\end{figure*}

	To analyse the baseline performance of the multiplexed sensor, we consider the ideal case where the aberrating wavefront consists entirely of modes to which the HMWS is sensitive. To demonstrate the interchangeability of the sensor mode basis, two distinct sensor designs are considered, which for ease of comparison both utilize six sensing modes each with bias $b_k=0.7$ and an APP dark hole of radial extent $2.7-6\lambda/D$. Sensor A encodes the first six non-trivial Zernike modes (Defocus $Z_3$ to Trefoil $Z_8$) while Sensor B contains six sinusoidal 2D Fourier modes of the form ${\rm{cos}}((n_xX+n_yY) + c)$, where $c$ is equal to either 0 or $\pi$, optimised to probe three critical locations at radial separation $3.5\lambda/D$ within the APP dark hole. The diffraction-limited PSFs of these sensors can be seen in Fig~\ref{Fig:CL_sims}b, with PSF copies showing the characteristic PSF of each sensing mode. Note that the APP of the Zernike cMWS is optimised for a $180^\circ$ opening angle while the Fourier cMWS contains an APP optimised for $90^\circ$, which explains the differences between the two diffraction-limited zeroth-order PSFs.
		
	Aberrating wavefronts are generated with equal RMS wavefront error $a$ present in each mode, giving a total RMS wavefront error $\sigma_\phi = \sum_k a_kM_k = a\sqrt{6}$ for a perfectly orthogonal 6-mode basis. In order to probe the upper limit of closed-loop convergence $a$ is varied between 0.1 and 1.5~radians RMS per mode, significantly exceeding the nominal $\pm0.5$~radians RMS per mode linear range of the sensor. The response matrix is constructed according to Eq.~\ref{Eq:S_ij} from a simulated calibration dataset, and compensation for the APP mode coefficients applied as per Eq.~\ref{Eq: APP_CTcorr}. Photometric apertures of radius $r_s = 1.22\lambda/D$ are applied to each PSF copy for flux measurement, which has been shown to provide optimal sensitivity for small $b_k$ \citep{Booth:03}. Closed-loop correction is then achieved by direct phase conjugation using a perfect simulated deformable mirror with phase $\Phi_{DM,i} = \Phi_{DM,i-1}-g\Sigma_k^N a_k M_k$, with the closed-loop gain $g$ left as a free parameter. Convergence is taken to be achieved at iteration $N_i$ where the total wavefront error $a_r$ is reduced below $10^{-2}$~radians RMS, which is seen to correspond closely to the point at which the diffraction-limited PSF is recovered.
		
	The panels of Fig~\ref{Fig:CL_sims} shows one example of closed-loop convergence for both sensors, with initial wavefront error of $a_k=1.0$~radians RMS per mode (and thus total wavefront error $\sigma_\phi = 2.45$~radians RMS) and a closed-loop gain $g=0.8$. It can be seen that despite this large initial wavefront error both sensors efficiently recover diffraction-limited APP performance with iteration number $N_i$, with residual wavefront error continuing to decline logarithmically towards the numerical noise threshold after nominal convergence is achieved. In this case the remaining intensity structure in the dark hole is limited purely by the HMWS scattered light background for each APP design. It is unclear exactly why the Fourier mode basis exhibits significantly faster convergence in this example, but a probable explanation is that the large coma aberration present as part of both APP designs pushes the Zernike mode sensor into the nonlinear regime and thus lowers the initial measurment accuracy of this mode, whereas this same aberration is distributed more evenly in the Fourier mode basis.
	
\begin{figure*}
		\centering
		\includegraphics[width=0.95\textwidth]{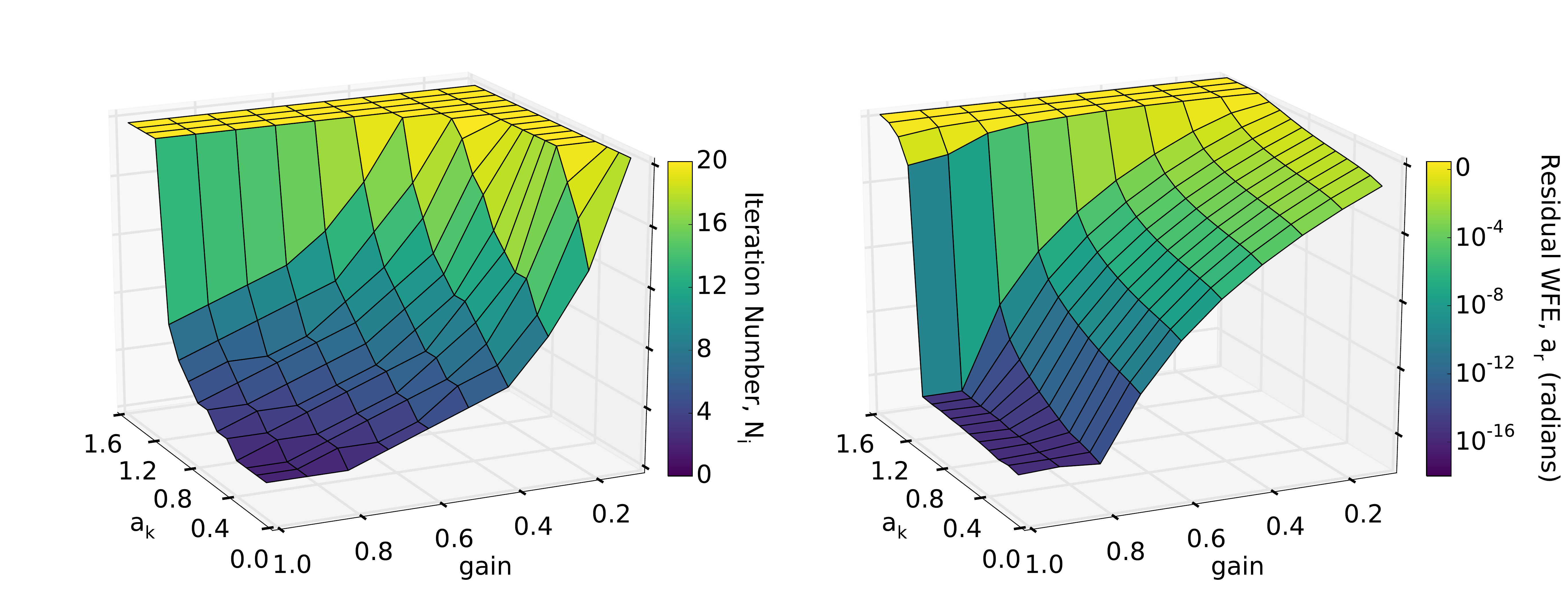}
		\caption{Convergence properties of the Zernike cMWS shown in Fig.~\ref{Fig:CL_sims} as a function of initial RMS wavefront error per mode, $a_k$, and closed-loop gain, $g$. {\bf Left:} Number of iterations required until convergence $N_i(a_r < 10^{-2})$. A value of $N_i=20$ indicates that convergence was not achieved within the allowed 20 iterations. {\bf Right:} Residual wavefront error $a_r$ after the final iteration. }
		\label{Fig:CL_surface}
		\end{figure*}

		Fig.~\ref{Fig:CL_surface} characterises in detail the convergence efficiency of the Zernike mode sensor by considering a wide variety of closed-loop gains $g$ and input RMS wavefront errors $a_r$. Both panels show that the critical failure point of this sensor lies at $a_k = 1.1$~radians RMS per mode and is independent of gain value. Below this, convergence speed is purely gain-limited for $g<0.8$ and $g=1$ provides the most efficient convergence for all $a_k$, ranging from $2 < N_i < 7$ iterations and with final $N_i = 20$ solutions consistent with the diffraction-limited wavefront at the level of numerical noise. This robust high-gain convergence behaviour stems from systematic underestimation of the wavefront outside the linear range (see Fig.~\ref{Fig:2c_Rcurve}), preventing oscillatory instabilities from occurring. The rapid breakdown in convergence above $a_k = 1.1$ happens when the contribution of nonlinear intermodal crosstalk between 6 modes of equal $a_k$ becomes comparable to the individual sensor response, enabling sign errors and thus irreversible divergence. The equivalent surface plots for the Fourier-type sensor was seen to be morphologically identical, confirming that the HMWS is capable of operating with any mode basis that is sufficiently complete with respect to the power spectrum of wavefront error present in the system.
		
		It is important to note that the term “idealised” here refers to the fact that no artificial noise sources such as readout or photon noise are included in these simulations, and that the underlying light source is purely monochromatic and point-like in nature. Such factors are dealt with during the on-sky implementation of the cMWS presented in Sec.~\ref{Sec: WHT Results} of this paper; in this section we instead aim to demonstrate that fundamental factors such as the multiplexing process and inter-modal crosstalk do not limit the final convergence of the closed-loop correction process. 
	This explains why the residual wavefront error as presented in Fig.~\ref{Fig:CL_sims} reaches the numerical noise limit in both examples; this will not be the case in practice as noise sources will result in sporadic random errors in measuring the wavefront coefficients. In the absence of systematic errors this can be expected to stall the convergence process at the level of $\sim 10^{-1}$ radians RMS based on the error bars derived in Sec.~\ref{Sec: WHT Results}, although this ultimately depends upon the signal-to-noise ratio (SNR) of individual WFS spots on a target-by-target basis. As presented in Sec.~\ref{Sec: BB_WFS}, use of a broadband source turns the holographic PSF copies into radially dispersed spectra, which can be useful in its own right for wavelength selection of the wavefront estimates.
		
\section{On-sky demonstration}
	\label{Sec: WHT Results}

	\subsection{Instrument design}
	\label{Sec: Expt}
		
\begin{figure}
		\centering
		\includegraphics[width=0.49\textwidth]{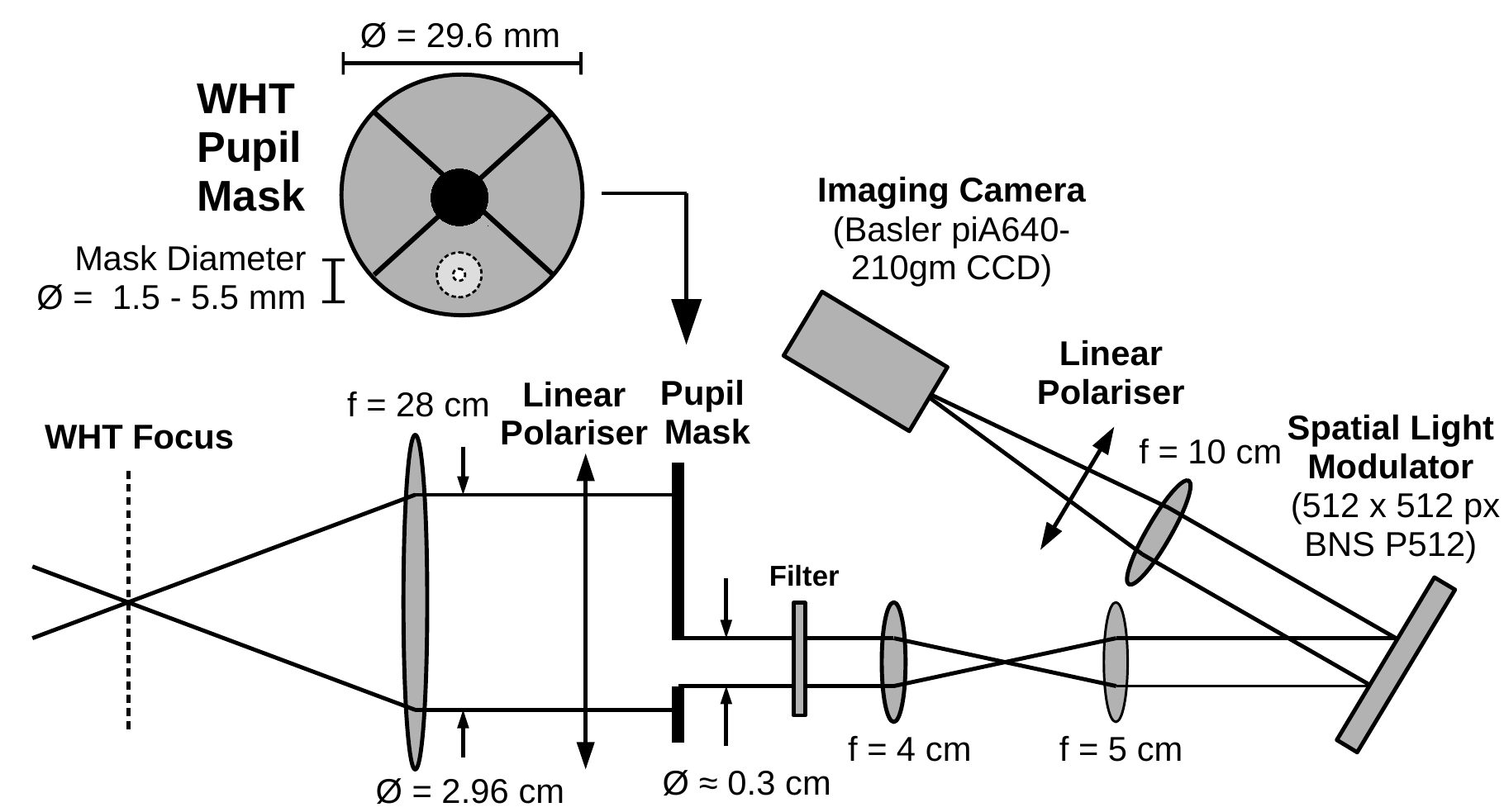}
		\caption{Diagram illustrating setup used at the WHT. Interchangeable pupil masks allowed the use of telescope sub-apertures with effective diameters ranging from 0.2~m to 0.8~m, to control relative aberration strengths in the absence of a classical AO system.}
		\label{Fig:4_SLMstetup}
	\end{figure}

	To implement the sensor on-sky at the William Herschel Telescope, we used a setup based around a BNS P512 reflective Spatial Light Modulator (SLM) as shown in Fig.~\ref{Fig:4_SLMstetup}, similar to that described in \cite{Kork:14}. This was operated with 250 pixels across the pupil diameter, oversampling the cMWS designs by a factor of two in order to ensure the sharp boundary regions of the HMWS hologram are accurately represented. Use of an SLM allows the rapid testing of a wide variety of designs without the need to manufacture individual custom optics, but has the disadvantage of allowing only passive measurement of wavefront errors: the response rate of the SLM was seen to approach 1Hz at times and as such is not a suitable active element for real-time phase correction.
	The SLM phase response was calibrated at the He-Ne 633~nm line via the differential optical transfer function (dOTF) wavefront reconstruction method of \cite{Kork:13}, at which the SLM is able to produce a maximum stroke of $1.94\pi$~radians. This stroke limitation to less than $2\pi$ is unimportant as all chosen designs have peak-to-peak phase values of less than $\pi$~radians. The sensor was then operated on-sky with both narrowband (650~nm, $\Delta\lambda = 10~nm$) and broadband (Bessel-R 550-900~nm) filters, with the latter possible despite strong chromatic behaviour of SLM devices (see Sec~\ref{Sec: BB_WFS} for further discussion). A high-frame rate Basler piA640-210gm CCD camera was used to record the focal plane including the holographic WFS spots at a cadence of 50Hz, comparable to atmospheric seeing timescales.
	
	It was necessary to limit on-sky wavefront error to within the dynamic range of the sensor, which in the absence of an AO system was achieved by stopping down the WHT aperture. For this purpose an off-axis circular pupil stop was used to create an un-obscured sub-aperture of effective diameter 42.3cm, positioned in the pupil so as to be free of telescope spiders over the elevation range $30\deg$ to zenith. This aperture size was chosen based on the expectation values of low-order Zernike coefficients of a pure Kolmogorov phase screen, which are constrained to $0.1 \lesssim \left|a_k\right| \lesssim 0.5 $~radians RMS for the 0.7"-2.5" range of seeing conditions typical of La Palma.
	
\begin{figure*}
		\centering
		\includegraphics[width=0.9\textwidth]{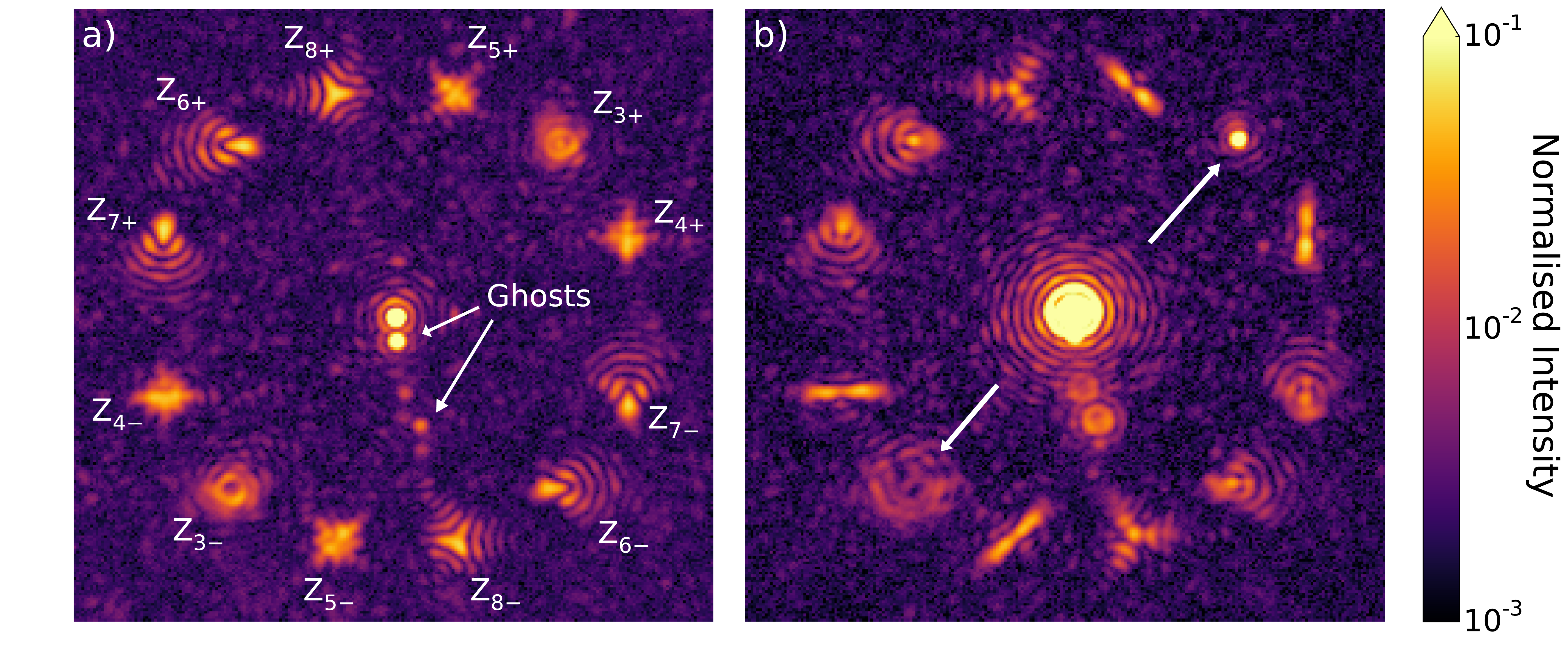}
		\caption{633~nm He-Ne laser calibration data illustrating the response of a 6-mode HMWS to controlled wavefront aberrations. {\rm a)} PSF with no induced wavefront error and {\rm b)} with 1.5~rad of defocus error applied via the SLM, showing the  asymmetric response of the $Z_3\pm$ defocus WFS spot pair, indicated top-right/bottom-left. Filter and SLM reflection ghosts are present below the zeroth-order PSF, though these are sufficiently separated from the PSF copies to ensure no interference with HMWS performance.}
		\label{Fig:5_Img_NB}
	\end{figure*}

	Two calibration images of a 6-mode Zernike HMWS with uniform bias value $b=1.5$~radians RMS at the calibration wavelength are shown in Fig~\ref{Fig:5_Img_NB}, for a flat wavefront and for 1.5~radians RMS of defocus error introduced on the SLM. For ease of illustration, a grating amplitude of $s=3\pi/4~\rm{radians}$ results here in an effective Strehl ratio of $24\%$ compared to the un-aberrated PSF. This illustrates clearly the sensor response: since no APP is applied in this instance, the holographic copy which is biased with a focus aberration of equal amplitude but opposite sign ($b_k = -a_k$) collapses to the Airy diffraction function, while the conjugate WFS spot gains double the aberration. 
	It should be noted that in addition to three faint filter ghosts below the zeroth order PSF, there is a significant ghost located at approximately $3\lambda/D$ which proved impossible to eliminate via optical re-alignment. This is attributed to unwanted reflection from the SLM glass cover plate which thus bypasses the active surface; a conclusion which is supported by its presence adjacent to both the central PSF and each filter ghost but not diffracted PSF copies, plus its independence of SLM-induced defocus.
		
	An in-situ calibration of the HMWS response matrix was obtained by sequentially introducing aberrations $a_kM_k$ with the SLM. It was found however that this solution contained linear inter-modal crosstalk components (off-diagonal terms in ${\bf \hat{G}}$) on the same order as sensor linear response. This effect is not seen in simulations nor indeed in un-calibrated normalised intensity data when the defocus error is generated with an external source (see the following section). The effect is therefore attributed to errors in accurately recreating Zernike modes with the SLM due to uncontrolled spatial variations in SLM voltage-to-phase actuation response across the pupil, which degrades the mode orthogonality. We therefore rely on simulated response matrix solutions for the following analysis of on-sky sensor data.
		
	\subsection{Characterising HMWS on-sky response}
	\label{Sec: WHT Defocus}
		
	To the authors' knowledge the HMWS has never before been implemented on-sky: the first and most important test was therefore to verify the on-sky response of the HMWS alone to known, static wavefront errors under realistic observing conditions. This is particularly important with respect to the ultimate goal of NCPE sensing, as the cMWS must be able to accurately recover coherent errors from underneath a dominant incoherent atmospheric speckle foreground.
	For this purpose, narrowband 650~nm observations were made of Arcturus ($m_R = -1.03$) using the HMWS design of Fig.~\ref{Fig:5_Img_NB} while scanning over a range of focus positions with the WHT secondary mirror, thereby inducing defocus error ranging between $\pm 2$~radians RMS in a controlled manner. 60 seconds worth of 20~ms exposures were stacked for each focus position in order to increase the signal-to-noise ratio and sufficiently average out atmospheric variations.
	The core intensity measurements of each WFS spot were then extracted using a numerical photometric aperture mask comprising a set of circular apertures of radius $1.22\lambda /D$, aligned with the centroid location of each WFS spot. The resulting set of intensity measurements was then converted to mode coefficient estimates $a_j$ using Eqs.~\ref{Eq:TypeB} and~\ref{Eq: APP_CTcorr}.
	
\begin{figure*}
		\centering
		\includegraphics[width=0.8\textwidth]{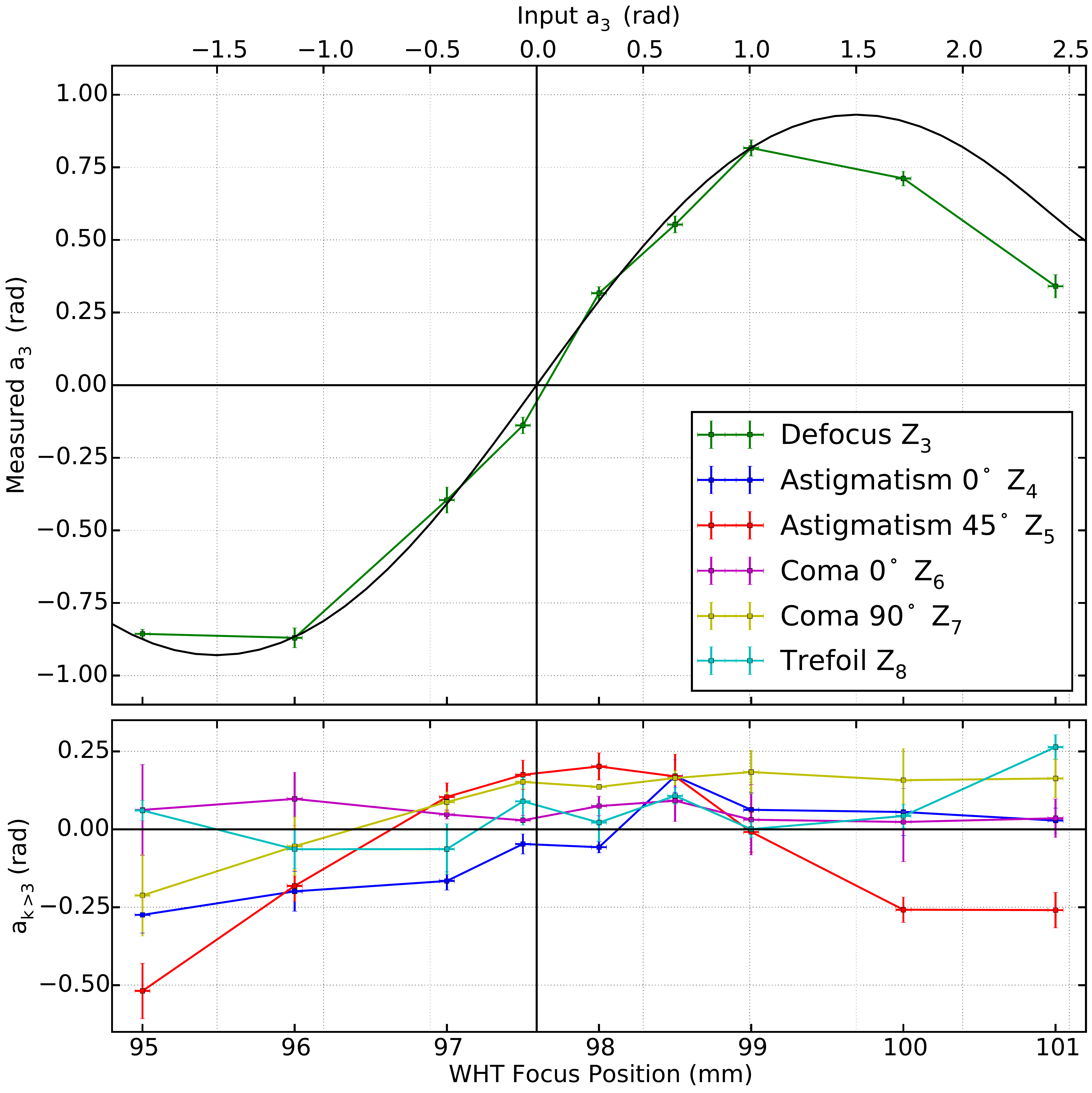}
		\caption{On-sky response of the Zernike HMWS sensor as a function of WHT secondary mirror focus position (mm) and corresponding induced defocus error $a_k$ (radians RMS). 
		{\bf Top:} Response of the $Z_3$ defocus mode. Overplotted in black is the theoretical $a_3$ response curve of Fig.~\ref{Fig:2c_Rcurve}, seen to correspond closely to sensor measurements over the the range $a_3 < 1.5$~radians RMS.
		{\bf Bottom:} Response of non-defocus modes $k>3$. Error bars correspond to uncertainties in sub-pixel centering accuracy (photon and CCD noise sources are negligible). Systematic trends in the non-defocus modes are attributed to real, instrumentally induced wavefront aberrations upstream of the pupil mask.}
		\label{Fig:6_StaticErr_sense}
	\end{figure*}

	Fig.~\ref{Fig:6_StaticErr_sense} shows the calibrated sensor response of all modes as a function of WHT focus position (mm), and the corresponding input defocus error $a_3$ in radians RMS. The mm-to-radians RMS scaling factor was obtained by least-squares fitting of the theoretical defocus response curve of Fig.~\ref{Fig:2c_Rcurve} (plotted here in black) to the defocus data, which is seen to be closely consistent for input $a_3 < 1.5$~radians RMS. It is unclear why the final two points are underestimated with respect to the theoretical curve, but even assuming this is a real effect, the additional wavefront underestimation in this nonlinear regime would have little impact on closed-loop sensor performance.
	The error bars on each curve represent the uncertainty in frame alignment, specifically the $1\sigma$ standard deviation of sensor measurements associated with the complete set of possible 1 pixel translational and rotational offsets of the photometric aperture mask, which was seen to be the practical limit on frame alignment accuracy. It was found that this source of uncertainty dominates over photon and readout noise when analysing seeing-averaged images; this places a limit on the precision of cMWS wavefront retrieval in the high-SNR regime of 0.04 radians per mode, or equivalently 0.1 radians RMS total wavefront error, a value obtained from the mean derived $1\sigma$ error bar of all six sensing modes where the input focus error is within the $a_3=\pm1$~radian RMS dynamic range of the sensor. 
	 Being azimuthally symmetric, the defocus mode is seen to be significantly more robust against small (x',y') offsets or rotations of the photometric aperture mask compared to other modes, even for large wavefront errors. Stability against positioning and/or tip-tilt errors is therefore a worthwhile consideration in choice of mode basis for future sensor designs.
	
	It can be seen that the response curves of the other sensed modes depart significantly from the well-behaved off-diagonal terms in Fig.~\ref{Fig:2c_Rcurve}; in particular the astigmatism mode $Z_5$ displays strongly quadratic response as a function of defocus error, which cannot be corrected using a linear interaction matrix and may therefore lead to closed-loop instabilities where large focus errors are present. This behaviour indicates that either the true sensor response is not fully characterised by the simulated interaction matrix, in which case experimental calibration is necessary, or that the injected wavefront error contains variable components other than pure defocus.
	This second hypothesis has been further explored because of the complexities associated with using the WHT secondary mirror as the source of injected focus error when we are sampling only an off-axis sub-pupil, as illustrated in the ``WHT Pupil Mask'' inset of Fig.~\ref{Fig:4_SLMstetup}. It was found in simulations that the observed crosstalk behaviour can be recreated if there are also static higher-order wavefront errors present upstream of the pupil mask, created for instance by small mis-alignments of the upstream lens or polariser. These aberrations are not orthogonal on the Zernike basis of the sensor and can mix with the variable focus error when sampled in this way, creating a spectrum of focus-dependent higher-order Zernike aberrations which cannot be explained with the modes $Z_{0-3}$ alone due to symmetry arguments.
	In this case it was found that adding an upstream wavefront error of 0.2~radians RMS of $Z_4$ (astigmatism of the opposite orientation) results in a $Z_5$ response curve which is morphologically similar to that shown in the lower panel of Fig.~\ref{Fig:6_StaticErr_sense}.  Low-amplitude variations seen in the coma and trefoil response curves are more difficult to recreate using static errors of the same order and may therefore be due to other factors, but it is expected that the principle remains the same when including additional Zernike mode orders $Z_{k>8}$.	
	By applying the correct set of upstream instrumental aberrations in this way it may be possible to account for the complete discrepancy between the non-defocus mode response of Fig.~\ref{Fig:2c_Rcurve} and the lower panel of Fig.~\ref{Fig:6_StaticErr_sense}. However, due to the complexity of this effect a comprehensive treatment is beyond the scope of this paper, which is in any case specific to the non-standard pupil apodisation used in this setup and is therefore not expected to be present in subsequent observing campaigns.
	
	It is sufficient to note that the defocus response is consistent with theoretical predictions, while the majority of the remaining sensor behaviour can be explained by unintentionally introduced additional instrumental wavefront errors and not to fundamental crosstalk effects, which appear to be accurately compensated by the theoretical response matrix calibration procedure. This confirms that the sensor is able to accurately recover (quasi-)static errors underneath a dominant fluctuating atmospheric speckle foreground, simply by integrating up to the desired timescale.
	
	\subsection{Broadband wavefront sensing}
	\label{Sec: BB_WFS}
	
	It is also important to characterise the broadband performance of the cMWS; a major limitation of focal-plane phase-retrieval algorithms such as phase diversity is that they only work effectively in the monochromatic case \citep{Kork:14}. By contrast, the HMWS contains no such fundamental limitations; the normalised difference metric is independent of variations in spectral transmission $T(\lambda)$, while the natural $\lambda$-dependence of the radial position of diffracted holographic PSF copies raises the intriguing possibility of performing wavelength-resolved wavefront sensing. 
	Spatial light modulators also typically exhibit strong chromatic response variations away from the calibration wavelength (e.g. \citealp{Spang:14}), but since all wavefront bias information is encoded into spatial variations in the binary hologram, only the effective grating amplitude $s$ and thus $T(\lambda)$ may vary with wavelength rather than bias $b_k$. 
	Altogether, the cMWS is in principle capable of delivering unbiased estimates of the wavefront coefficients $a_k$ in radians RMS for arbitrarily wide spectral bands, at a spectral resolution set by the diffraction limit of the monochromatic telescope PSF.
		
\begin{figure*}
		\centering
		\includegraphics[width=\textwidth]{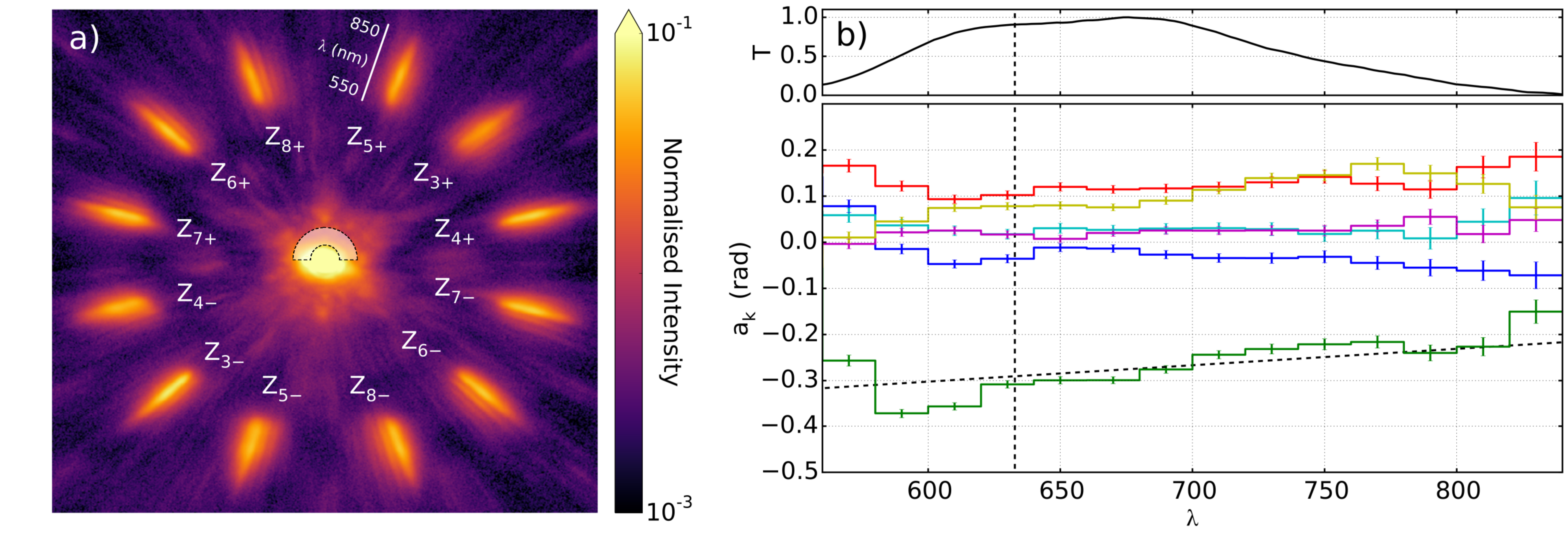}
		\caption{\textbf{a)} Broadband on-sky PSF of a cMWS incorporating the HMWS of Fig.~\ref{Fig:5_Img_NB}, showing $1/\lambda$ radial dispersion of PSF copies. The location of the APP dark hole at $\lambda=633$~nm is illustrated by the shaded region, spanning $2.7-6\lambda/D$, although this is not directly visible in the stacked image data. \textbf{b)} Chromatic response of each mode, binned to $\Delta \lambda = 20$~nm wavelength intervals; colours correspond to the modes of Fig.~\ref{Fig:6_StaticErr_sense}.  The 633~nm calibration wavelength is shown by the vertical dashed line, and the mean spectral transmission $T(\lambda)$ is shown in the upper panel. Defocus $Z_3$ (green) shows indications of $1/\lambda$ scaling, illustrated by the diagonal dashed line. }
		\label{Fig:7_BBimg}
	\end{figure*}

	Fig.~\ref{Fig:7_BBimg}a shows the broadband on-sky PSF of a cMWS including the same 6-mode Zernike HMWS as in Fig.~\ref{Fig:5_Img_NB}, operated with a standard $50\%$ bandwidth Bessel-R filter. To test the full cMWS concept, this design also includes a $180$ degree APP as in Fig.~\ref{Fig:2_HMWS_simImg}; in this seeing-limited image the dark hole is located in the top half of the PSF, although it is obscured by residual speckles and chromatic dispersion. In Fig.~\ref{Fig:7_BBimg}b it can be seen that the chromatic response of each mode is broadly consistent over the FWHM transmission range of 580-750~nm. A residual focus error can be clearly seen from these on-sky observations, such that this plot corresponds to the wavelength dimension of $a_3 \approx -0.3$~rad in Fig.~\ref{Fig:6_StaticErr_sense}. This mode also displays the $a_k \propto 1/\lambda$ scaling expected from physical wavefront errors.

	Confirmation of a bias-free spectral response allows boosting of single-frame SNR by binning the 580-750~nm spectrum in the radial and hence wavelength dimension, making quasi-real-time wavefront sensing, with exposure times $t_{exp}$ approximately equal to the NCPE coherenct timescale $\tau_\phi$, a possibility. Wavelength-resolved wavefront sensing may also be achieved by using appropriately calibrated photometric sub-apertures along the dispersed wavefront spectra, and will be considered further in future work. Such information may be useful for optimisation of the broadband control of existing AO systems or for next-generation instrument concepts consisting of of multiple corrective elements for specific wavelength ranges.
	
	\subsection{Real-time atmospheric wavefront measurements}
	
\begin{figure*}
		\centering
		\includegraphics[width=\textwidth]{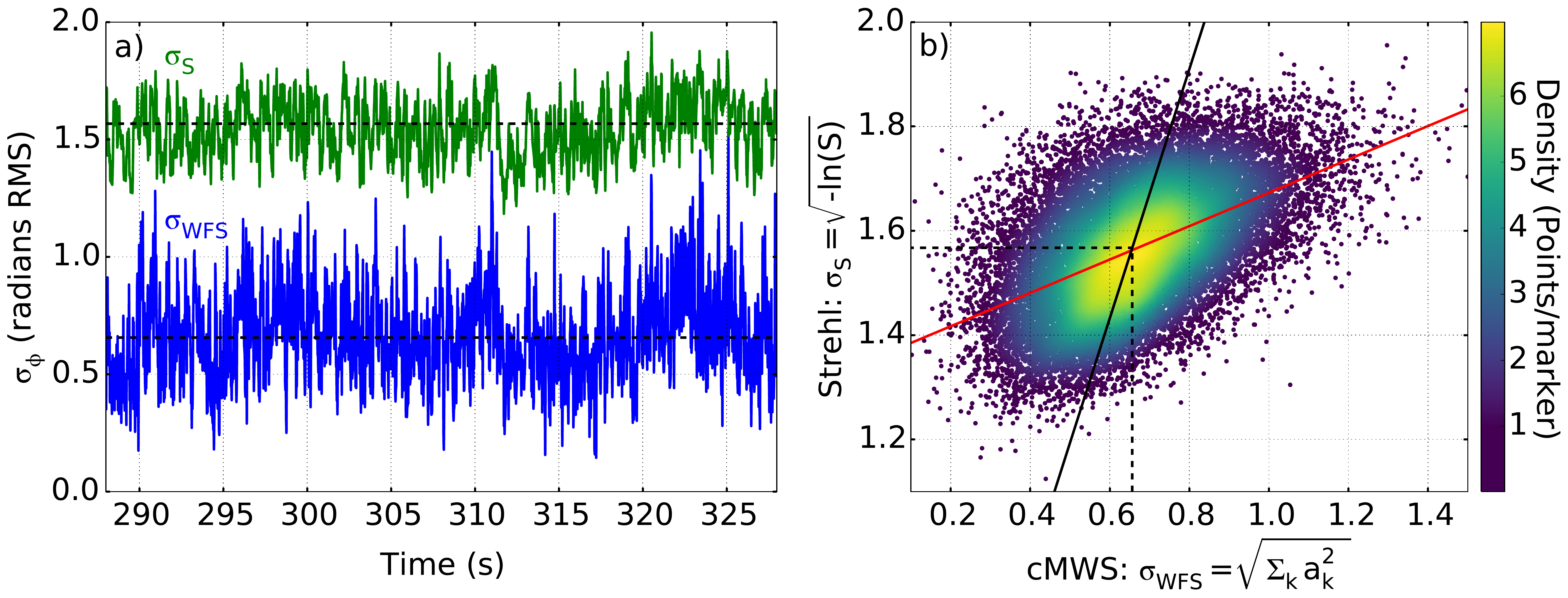}
		\caption{{\bf a)} Time series of single-frame RMS wavefront error $\sigma_\phi$ as measured by i) the central PSF Strehl ratio under the Mar\'{e}chal approximation ($\sigma_{\rm S}$, green) and ii) calibrated cMWS measurements ($\sigma_{\rm WFS}$, blue). Slow variations in seeing quality seen in $\sigma_{\rm S}$ are visibly traced by the sensor. {\bf b)} Correlation plot between the two independent estimates of RMS wavefront error, with colour indicating point density. The solid black line denotes $\sigma_{\rm S}/\sigma_{\rm WFS} = 0.45$ as is expected theoretically from \cite{Noll:76}, which corresponds well to the core regions of the correlation ($1.45 < \sigma_{\rm S} < 1.65$). The outer regions have a significantly shallower gradient (red), distorted by various systematic error sources.}
		\label{Fig:9_AtmosSense}
	\end{figure*}

	Application of the broadband sensor provided sufficient SNR for partial wavefront retrieval from individual 20ms frames. Fig.~\ref{Fig:9_AtmosSense}a shows two independent estimators of RMS wavefront error $\sigma_\phi$ for a subset of frames: cMWS measurements $\sigma_{\phi,cMWS} = \sqrt{\Sigma_k {a_k}^2}$ (lower, blue curve) and $\sigma_{\phi,S} = \sqrt{-{\rm ln}(S)}$ from Strehl ratio measurements of the science PSF under the Mar\'{e}chal approximation (upper, green curve). Despite significant noise in the measurements, the cMWS measurements trace slow trends in image Strehl ratio on timescales $>1$~s, associated with changing seeing conditions. 
	Fig.~\ref{Fig:9_AtmosSense}b shows the resulting correlation between these two frame quality estimators for the full 20,790 frame dataset spanning 10 minutes of observation, with a Pearson correlation index of $\rho = 0.50$. Frames with $\left|I_k\right|>1$ were rejected as such measurements are obviously unphysical: such events are rare ($<1\%$ of total frames affected) and attributed to cosmic ray impacts and residual hot/cold pixels. 
	Additional confirmation that the cMWS is tracing the atmospheric wavefront is provided by the respective mean wavefront error estimates: $\langle\sigma_{\phi,cMWS}\rangle = 0.656 \pm 0.001$ and $\langle\sigma_{\phi,S}\rangle = 1.567 \pm 0.001$~radians RMS, such that on average the cMWS senses approximately $42\%$ of the total wavefront error once known static errors have been subtracted. This is notably consistent with what is expected for pure Kolmogorov statistics as presented in \cite{Noll:76}, where $45\%$ of the total wavefront error for $Z_{k>3}$ is contained in the first 6 non-trivial modes; this relation $\sigma_{\rm S}/\sigma_{\rm WFS} = 0.45$ is denoted by the solid black line in Fig.~\ref{Fig:9_AtmosSense}b.	
	Ideally the two independent estimates should correlate along this relation for all values of wavefront error, but although there is reasonable agreement about the mean $\sigma_{\rm WFS} = 0.65$~radians RMS, it can be seen that the correlation is significantly shallower for outlying points beyond $\sigma_{\rm WFS}>0.8$~radians RMS; here the best fit line, plotted in red, clearly does not intersect the origin. This may be attributed to crosstalk with high-order unsensed modes allowing the sensor to pick up some additional wavefront error to that contained purely in the 6-mode basis, or to systematic effects such as sensor saturation, making these extreme wavefront estimates unreliable. However, it is important that the majority of sensor measurement points fall close to the theoretically expected relation, where sensor performance is expected to be most reliable.
		
\begin{figure}
		\centering
		\includegraphics[width=0.45\textwidth]{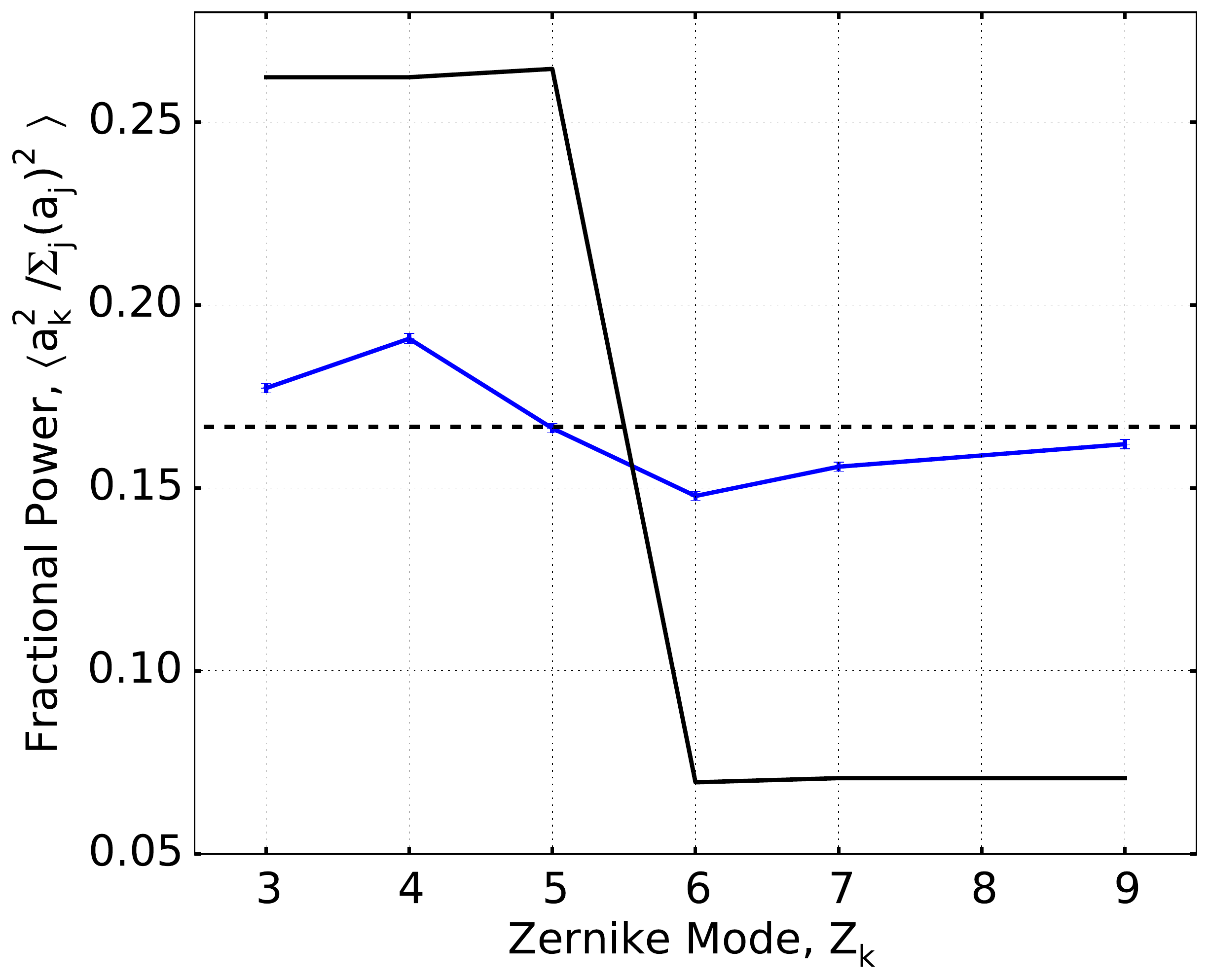}
		\caption{Modal power spectrum of on-sky broadband cMWS measurements as a function of Zernike mode order. The theoretical spectrum corresponding to purely Kolmogorov statistics \citep{Noll:76} is overplotted in black, while the horizontal dashed line denotes a purely flat 6-mode power spectrum.}
		\label{Fig:10_MPS}
	\end{figure}

	The ultimate goal of this process is to reconstruct the instantaneous wavefront in each frame. As may be anticipated from Fig.~\ref{Fig:9_AtmosSense}a however, such wavefronts were seen to be dominated by frame-to-frame noise. In order to assess the extent to which the independent mode coefficient measurements are degraded, we plot the modal power spectrum of the full dataset in Fig.~\ref{Fig:10_MPS} and contrast with that expected from Kolmogorov turbulence as rescaled to a six-mode basis.
	It can be seen that although there is some morphological similarity which indicates a decreasing power spectrum, the amplitude of individual modes is significantly more consistent with a flat spectrum. It is possible that the true seeing statistics are not Kolmogorov in nature, however it is difficult to justify a discrepancy of such size in this manner. Instead, it is assumed that this is due to the mixing effect of crosstalk with higher-order un-sensed modes which cannot be accounted for by the response matrix; only in this way is it possible to preserve the total wavefront variance as discussed above. The immediate solution for residual atmospheric wavefront error sensing is to increase the number of modes to encompass a larger fraction of the total power spectrum. For the application to NCPE correction of a dark hole the problem is made simpler as the power spectrum is expected to be dominated by low-order components, which may be accessed by integrating so as to sufficiently average out the unwanted high-order atmospheric errors.
	
	Due to the dominance of frame-to-frame noise at a cadence of 50Hz we therefore draw only limited conclusions regarding the potential of the cMWS for real-time wavefront correction in this instance, however the successful retrieval of total wavefront error $\sigma_\phi$ at this cadence is already a promising result for such a preliminary test. More important is that, as shown in Fig.~\ref{Fig:6_StaticErr_sense}, the cMWS is capable of recovering known static aberrations to a precision of approximately 0.04 radians RMS per mode with one-shot measurements. This is performed in the presence of a dominant and fluctuating atmospheric speckle foreground, in direct parallel with the ultimate goal of direct NCPE sensing.

\section{Discussion \& conclusions}
\label{Sec: Conclusion}

We have demonstrated via idealised closed-loop simulations and a first on-sky implementation that the coronagraphic Modal Wavefront Sensor (cMWS) is a promising new focal-plane sensor for high-contrast imaging, highly suited to correction of  Non-Common Path Errors (NCPEs) and with additional potential as a high-cadence broadband wavefront sensor. 
The major advantage of the cMWS over prior focal-plane reconstruction algorithms is that the measurement process requires no invasive modification of the science PSF, as is required for phase diversity approaches; this allows the correction loop to be effectively decoupled from science observations.

The performance of the cMWS is not limited by the process of multiplexing the APP coronagraph and Holographic Modal Wavefront Sensor (HMWS) constituents or structures in the telescope aperture function, but by residual inter-modal crosstalk with higher-order unsensed modes present in the wavefront. This can be addressed by using a larger sensing mode basis than the 6-mode cMWS prototype presented in this work, such that a larger fraction of the total wavefront error is encompassed by the sensor.
The correction order of the cMWS is currently limited not by fundamental factors, but by the practical consideration of science PSF throughput. This may be optimised with respect to the signal-to-noise ratio of the holographic PSF copies and hence observational target brightness, however the practical limit in most cases is expected to be 20-30 modes. While expected to be sufficient for NCPE correction, this is too small to allow the removal of a classical AO sensor from the instrument design.
It is however possible to avoid such limitations for applications which require only a small but extremely well-corrected field of view, such as spectroscopic characterisation of known exoplanets. We have already discovered that it is possible to manufacture APP coronagraphs which reach simulated contrasts of $10^{-10}$ in dark regions a few square $\lambda/D$ in size (Keller et al., in prep.). These regions contain few degrees of freedom in the electric field such that they may be fully corrected with only a small basis of optimised modes.

An additional advantage of the cMWS is its computational simplicity, requiring only the relative photometry of the diffraction cores of 2$ N_{\rm mode}$ holographic PSF copies and a small number of linear computations for the calibration process; most importantly it does not require any Fourier transforms. This is unimportant for NCPE sensing due to the slow timescales involved, but an additional application of the sensor is then to the challenge of extremely high-cadence sensing, for the control of a limited number modes at kHz frequencies. Such an approach is expected to lead to significant improvements in wavefront quality over conventional AO update frequencies (Keller et. al., in prep.).
As a phase-only sensor, the on-sky performance of the cMWS will always be fundamentally limited by instrumental amplitude errors. This may be overcome by combining it with other focal-plane sensing techniques, such as electric field conjugation \cite{Giveon:06}, which are capable of reconstructing the full electric field but which lack the dynamic range to perform effectively by themselves in ground-based AO systems. The improved "Fast and Furious" algorithm of \cite{Kork:14} also lends itself to use with the cMWS, which naturally provides a large number of known phase diversities in the holographic PSF copies.

	Future work will focus on implementing the optimised cMWS behind a 97-actuator AO system with a classical Shack-Hartmann WFS, previously used with the ExPo high-contrast imaging polarimeter \citep{Roden:11}. In addition to providing a significant boost in SNR, this will allow the cMWS to be tested in a realistic closed-loop environment which reflects the ultimate goal of real-time NCPE control. If successful, such a system would be ideal for inclusion into the next-generation of high-contrast imaging instruments such as EPICS for the E-ELT \citep{Kasper:10}, for the detection and characterisation of rocky exoplanets in the habitable zones of nearby stars.
	
\begin{acknowledgements}
The authors would like to thank the anonymous referee, whose comments have helped to significantly improve this paper. Our thanks also to F. Riddick and the WHT operations team for assistance during on-sky implementation of the cMWS. This research is funded by the Nederlandse Onderzoekschool Voor Astronomie (NOVA).
\end{acknowledgements}

\bibliographystyle{aa}

\bibliography{Wilby16_cMWS}

\end{document}